\documentstyle[12pt,epsf]{article}
\newcommand{\gtap}{\;{\raise.3ex\hbox{$>$\kern-.75em\lower1ex\hbox{$\sim$}}}\;}
\newcommand{\ltap}{\;{\raise.3ex\hbox{$<$\kern-.75em\lower1ex\hbox{$\sim$}}}\;}

\newcommand{\D}[2]{\delta^{#1}_{#2}}

\begin{document}

\begin{titlepage}

\hspace*{\fill}\parbox[t]{6cm}
{hep-ph/0209271 \\ \\
ILL-(TH)-02-7 \\
FERMILAB-Pub-02/197-T \\ \\
\today} \vskip2cm
\begin{center}
{\Large \bf Color-flow decomposition of QCD amplitudes} \\
\medskip
\bigskip\bigskip\bigskip
{\large\bf F.~Maltoni,$^1$ K.~Paul,$^1$ T.~Stelzer,$^1$
S.~Willenbrock$^{1,2}$} \\
\bigskip\bigskip
$^1$Department of Physics \\
University of Illinois at Urbana-Champaign \\ 1110 West Green Street \\
Urbana, IL\ \ 61801 \\
\bigskip\bigskip
$^2$Theoretical Physics Department \\ Fermi National Accelerator Laboratory \\
P.~O.~Box 500 \\ Batavia, IL\ \ 60510 \\
\end{center}
\bigskip\bigskip\bigskip

\vspace{.5cm}

\begin{abstract}
We introduce a new color decomposition for multi-parton amplitudes in QCD,
free of fundamental-representation matrices and structure constants.  This
decomposition has a physical interpretation in terms of the flow of color,
which makes it ideal for merging with shower Monte-Carlo programs.  The
color-flow decomposition allows for very efficient evaluation of amplitudes
with many quarks and gluons, many times faster than the standard color
decomposition based on fundamental-representation matrices. This will increase
the speed of event generators for multi-jet processes, which are the principal
backgrounds to signals of new physics at colliders.
\end{abstract}

\vfil

\end{titlepage}

\section{Introduction}

Amplitudes with many external quarks and gluons are important for calculating
the cross section for multi-jet production (alone or together with other
particles) at the Fermilab Tevatron, CERN Large Hadron Collider, and future
$e^+e^-$ linear colliders. These processes are the major backgrounds to many
new-physics signals, so an accurate description of these final states is
essential.

Amplitudes involving many quarks and gluons are difficult to calculate, even
at tree level.  Over the years, techniques have been developed to calculate
these multi-parton amplitudes efficiently \cite{Mangano:1990by}.  One aspect of
such techniques is the systematic organization of the $SU(N)$ color
algebra.\footnote{Although we are interested specifically in QCD, for which
$N=3$, we leave $N$ unspecified whenever possible.} For example, consider the
amplitude for $n$ gluons of colors $a_1, a_2, \ldots, a_n$ ($a_i = 1, \ldots,
N^2 - 1$).  At tree level, such an amplitude can be decomposed as
\cite{Mangano:1987xk}
\begin{equation}
\label{eq:fund_decomposition} {\cal M}(ng) = \sum_{P(2,\ldots, n)} {\rm Tr}\,
(\lambda^{a_1} \lambda^{a_2} \cdots \lambda^{a_n})\; A(1,2,\ldots,n)\;,
\end{equation}
where $\lambda^{a}$ are the fundamental-representation matrices of $SU(N)$,
and the sum is over all $(n-1)!$ permutations of $(2,\ldots,n)$.  Each trace
corresponds to a particular color structure.  The factor associated with each
color structure, $A$, is called a partial amplitude.\footnote{Also referred to
as a dual amplitude or a color-ordered amplitude.} It depends on the
four-momenta $p_i$ and polarization vectors $\epsilon_i$ of the $n$ gluons,
represented simply by $i$ in the argument of the partial amplitude. These
partial amplitudes are far simpler to calculate than the full amplitude,
${\cal M}$, and they are also gauge invariant.  There exist linear relations
amongst the partial amplitudes, called Kleiss-Kuijf relations, which reduce
the number of linearly-independent partial amplitudes to $(n-2)!$
\cite{Kleiss:1988ne}.  A similar decomposition exists for amplitudes
containing any number of $\bar qq$ pairs and gluons.

Recently, another decomposition of the multi-gluon amplitude has been
introduced, based on the adjoint representation of $SU(N)$ rather than the
fundamental representation \cite{DelDuca:1999ha,DelDuca:1999rs}.  The
$n$-gluon amplitude in this decomposition may be written as
\begin{equation}
\label{eq:adjoint_decomposition} {\cal M}(ng) = \sum_{P(2,\ldots, n-1)}
(F^{a_2} F^{a_3} \cdots F^{a_{n-1}})^{a_1}_{a_n} \; A(1,2,\ldots,n)\;,
\end{equation}
where $(F^a)^b_c = -if^{a b c}$ are the adjoint-representation matrices of
$SU(N)$ ($f^{abc}$ are the structure constants), and the sum is over all
$(n-2)!$ permutations of $(2,\ldots,n-1)$.  The partial amplitudes that appear
in this decomposition are the same as in the other decomposition, but only the
$(n-2)!$ linearly-independent amplitudes are needed. The adjoint-representation
decomposition exists only for the multi-gluon amplitude.\footnote{There is yet
another decomposition of the $n$-gluon amplitude based on the adjoint
representation \cite{Berends:cv},
\begin{displaymath}
{\cal M}(ng) = \frac{1}{2N}\sum_{P(2,\ldots, n)} {\rm Tr}\, (F^{a_1} F^{a_2}
\cdots F^{a_n}) \; A(1,2,\ldots,n)\;,
\end{displaymath}
where the sum is over all $(n-1)!$ permutations of $(2,\ldots,n)$.  This was
the original color decomposition; it is no longer widely used.}

In this paper we introduce a third color decomposition of multi-parton
amplitudes.  This decomposition is based on treating the $SU(N)$ gluon field
as an $N \times N$ matrix $(A_{\mu})^{i}_{j}$ ($i,j = 1,\ldots, N$), rather
than as a one-index field $A_{\mu}^{a}$ ($a = 1,\ldots, N^2-1$).  The $n$-gluon
amplitude may be decomposed as
\begin{equation}
\label{eq:colorflow_decomposition} {\cal M}(ng) = \sum_{P(2,\ldots, n)}
\D{i_1}{j_2} \D{i_2}{j_3} \cdots \D{i_n}{j_1} \; A(1,2,\ldots,n)\;,
\end{equation}
where the sum is over all $(n-1)!$ permutations of $(2,\ldots,n)$.  We dub this
the {\it color-flow} decomposition, due to its physical interpretation.  The
partial amplitudes that appear in this decomposition are the same as in the
other two decompositions.  The proof of this assertion is contained in an
Appendix.

The color-flow decomposition has several nice features which we elaborate upon
in this paper.  First, a similar decomposition exists for all multi-parton
amplitudes, like the fundamental-representation decomposition.  Second, the
color-flow decomposition allows for a very efficient calculation of
multi-parton amplitudes.  For example, we show that the amplitude for 12
gluons ($gg\to 10g$) may be calculated about 60 times faster using the
color-flow decomposition than using the fundamental-representation
decomposition.  Third, it is a very natural way to decompose a QCD amplitude.
As the name suggests, it is based on the flow of color, so the decomposition
has a simple physical interpretation. This is also useful for merging the
hard-scattering cross section with shower Monte-Carlo programs.

The remainder of the paper is organized as follows.  In
Section~\ref{sec:Feynman} we derive the color-flow Feynman rules for the
construction of the partial amplitudes.  Section~\ref{sec:ngluon} is devoted
to the all-gluon amplitude.  In Section~\ref{sec:1quark} we consider the
amplitude for a $\bar qq$ pair and any number of gluons.
Section~\ref{sec:2quark} deals with the case of two $\bar qq$ pairs and any
number of gluons.  The general case is discussed in
Section~\ref{sec:general}.  We draw conclusions in
Section~\ref{sec:conclusions}.

\section{Feynman rules}\label{sec:Feynman}

Consider the Lagrangian of an $SU(N)$ gauge theory,
\begin{equation}
{\cal L} = \frac{1}{2 g^2} {\rm Tr}\,{F^{\mu \nu} F_{\mu \nu}} + \bar{\psi} (i
\not\!\! D - m) \psi\;,
\end{equation}
where
\begin{eqnarray}
D_{\mu} & = & \partial_{\mu} + i g A_{\mu} \\
F_{\mu \nu} & = & [ D_{\mu}, D_{\nu} ]\;.
\end{eqnarray}
The quark field transforms under the fundamental representation of $SU(N)$,
\begin{equation}
\psi \to U \psi\;.
\end{equation}
The gluon field $A_{\mu}$ transforms under the adjoint representation,
\begin{equation}
A_{\mu} \to U A_{\mu} U^{\dag} + \frac{i}{g}(\partial_{\mu} U) U^{\dag}\;,
\end{equation}
such that the Lagrangian is invariant under local $SU(N)$ transformations,
$U(x)$.

It is conventional to decompose the gluon field using the
fundamental-representation matrices,\footnote{${\rm Tr}\, (\lambda^{a}
\lambda^{b})=\delta^{ab}$}
\begin{equation}
(A_{\mu})^{i}_{j} = \frac{1}{\sqrt 2}A_{\mu}^{a} (\lambda^{a})^{i}_{j}\;,
\end{equation}
and to rewrite the Lagrangian in terms of $A_{\mu}^{a}$ ($a = 1, \ldots,
N^2-1$).  This yields the usual Feynman rules involving the
fundamental-representation matrices $\lambda^{a}$ and the structure constants
$f^{a b c}$, which arise from the commutation relation $[ \lambda^{a},
\lambda^{b} ] = i f^{a b c} \lambda^{c}$.  The decomposition of the $n$-gluon
amplitude in terms of traces of fundamental-representation matrices,
Eq.~(\ref{eq:fund_decomposition}), is then achieved by inverting the
commutation relation, $f^{a b c} = -i {\rm Tr}\, (\lambda^{a} \lambda^{b}
\lambda^{c} - \lambda^{a} \lambda^{c} \lambda^{b})$.

\begin{figure}[p]
\begin{center}
\vspace*{0cm} \hspace*{0cm} \epsfxsize=14cm \epsfbox{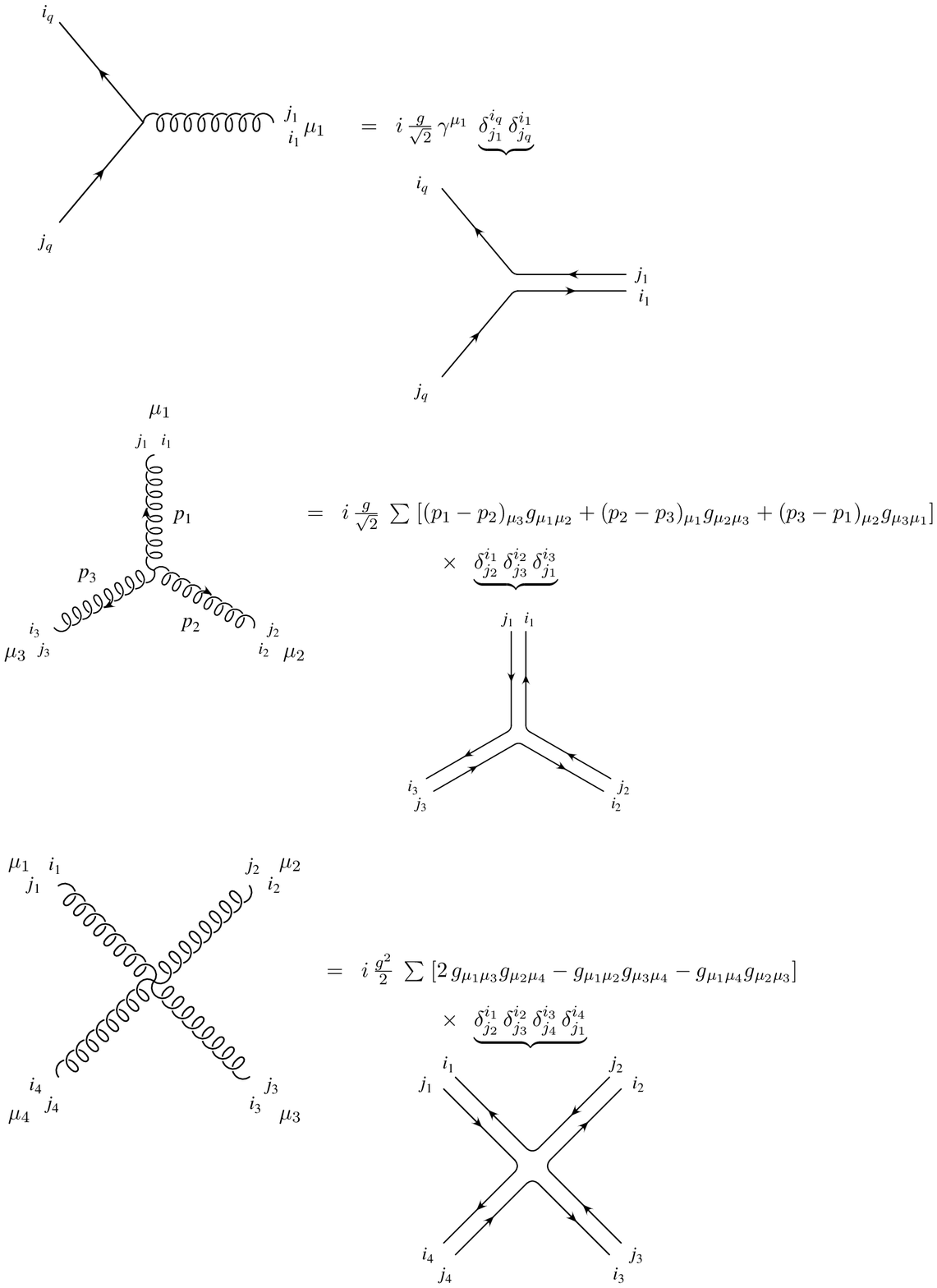}
\vspace*{0cm}
\end{center}
\caption{Color-flow Feynman rules.  All momenta are outgoing.  The
arrows indicate the flow of color.  The sum in the three-gluon
vertex is over the two non-cyclic permutations of (1,2,3); in the
four-gluon vertex, the sum is over the six non-cyclic permutations
of (1,2,3,4).  When calculating a partial amplitude the sum is
dropped, as only one term in the sum contributes to a given color
flow.} \label{fig:colorflow_feynmanrules}
\end{figure}

However, it is not necessary to decompose the gluon field $(A_{\mu})^{i}_{j}$
in terms of fundamental-representation matrices.  Instead, one can work
directly with the $N \times N$ matrix field $({\cal A}_{\mu})^{i}_{j} \equiv
\sqrt{2} ( A_{\mu})^{i}_{j}$.\footnote{The factor $\sqrt 2$ is introduced such
that the field is canonically normalized.} The Lagrangian is
\begin{equation}
{\cal L} = -\frac{1}{4}({\cal F}^{\mu \nu})^{i}_{j} ({\cal F}_{\mu
\nu})^{j}_{i} + i \bar{\psi}_{i} \gamma^\mu ( \delta^{i}_{j} \partial_{\mu} + i
\frac{g}{\sqrt{2}} ({\cal A}_{\mu})^{i}_{j} ) \psi^{j} - m \bar{\psi}_{i}
\psi^{i}\;,
\end{equation}
where
\begin{equation}
({\cal F}_{\mu \nu})^{i}_{j} = \partial_{\mu} ({\cal A}_{\nu})^{i}_{j} -
\partial_{\nu} ({\cal A}_{\mu})^{i}_{j} + i \frac{g}{\sqrt{2}}
({\cal A}_{\mu})^{i}_{k}({\cal A}_{\nu})^{k}_{j} -i \frac{g}{\sqrt{2}} ({\cal
A}_{\nu})^{i}_{k}({\cal A}_{\mu})^{k}_{j}\;.
\end{equation}
This yields Feynman rules free of fundamental-representation matrices and
structure constants.  These Feynman rules are given in
Fig.~\ref{fig:colorflow_feynmanrules}.\footnote{It is evident from the Feynman
rules that the natural coupling constant is $g/\sqrt 2$ rather than $g$.} This
representation of an $SU(N)$ gauge theory is well known from the $1/N$
expansion \cite{'tHooft:1973jz}.  However, it is not commonly used for
ordinary calculations in QCD.

In our conventions, upper indices transform under the fundamental
representation of $SU(N)$, lower indices under the antifundamental
representation.  Global $SU(N)$ symmetry implies that color is conserved at
the interaction vertices, just as electric charge is conserved at the
interaction vertex of QED.  Thus the interaction vertices may be represented by
color-flow Feynman rules, as shown in Fig.~\ref{fig:colorflow_feynmanrules}
\cite{Cvitanovic:1982rf}. The arrows track the flow of color from lower indices
to upper indices.  The three-gluon vertex has two color flows, and the
four-gluon vertex has six.

The gluon propagator is proportional to
\begin{equation}
\langle ({\cal A}_{\mu})^{i_1}_{j_1} ({\cal A}_{\nu})^{i_2}_{j_2} \rangle
\propto \D{i_1}{j_2} \D{i_2}{j_1} - \frac{1}{N} \D{i_1}{j_1} \D{i_2}{j_2}
\end{equation}
and thus has two different color flows. In contrast, the gluon propagator in
the conventional representation of color is proportional to
\begin{equation}
\langle A_{\mu}^{a} A_{\nu}^{b} \rangle \propto \delta^{a b} \;.
\end{equation}
The more complicated color structure of the gluon propagator is a trade-off for
the simplicity of the color structure of the interaction vertices
(Fig.~\ref{fig:colorflow_feynmanrules}).  As we shall see, this trade-off is
worthwhile.

\begin{figure}[t]
\begin{center}
\vspace*{0cm} \hspace*{0cm} \epsfxsize=8cm \epsfbox{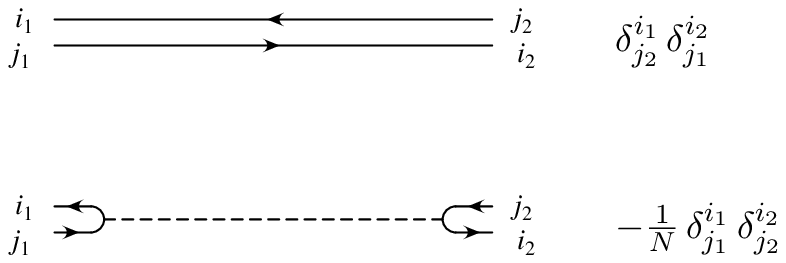}
\vspace*{0cm}
\end{center}
\caption{The $SU(N)$ gluon propagator may be split into a $U(N)$
gluon propagator and a $U(1)$ gluon propagator. The $U(1)$ gluon
interacts only with quarks.}
\label{fig:colorflow_gluon_propagator}
\end{figure}

Due to the antisymmetry of the three- and four-gluon vertices, the second
color flow in the gluon propagator does not couple to these interactions.  It
couples only to the gluon interaction with the quarks.  This color flow acts
as a ``photon'' that couples with strength $\frac{g}{\sqrt 2}$ to quarks.  We
indicate this by splitting the gluon propagator into a $U(N)$ gluon propagator
and a $U(1)$ gluon propagator, as shown in
Fig.~\ref{fig:colorflow_gluon_propagator}.  The $U(1)$ gluon is unphysical, as
evidenced by its ghostly residue, which also carries a factor $1/N$.

\section{$n$-gluon amplitude}\label{sec:ngluon}

The $n$-gluon tree-level amplitude is constructed from three- and four-gluon
vertices and the $U(N)$ gluon propagator; the $U(1)$ gluon propagator does not
couple to these interactions.  It follows directly from the color-flow Feynman
rules that the $n$-gluon tree-level amplitude has the decomposition given in
Eq.~(\ref{eq:colorflow_decomposition}).  We now describe how to calculate the
partial amplitude, which is the factor associated with a particular color flow.

\begin{figure}[b]
\begin{center}
\hspace*{0in} \epsfxsize=1.8in \epsfbox{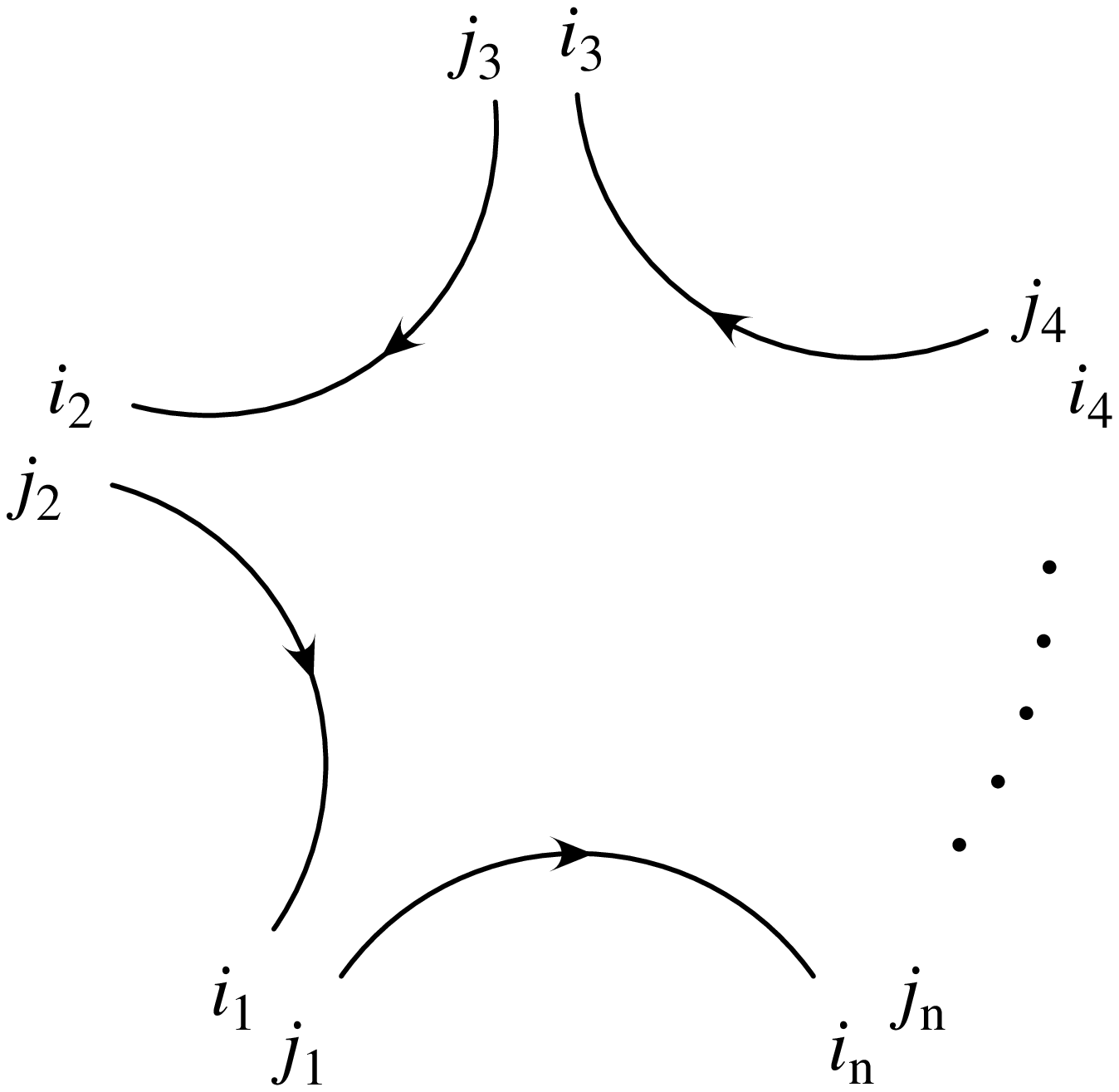}
\end{center}
\vspace*{-.1in} \caption{Color flow $\delta^{i_1}_{j_2} \delta^{i_2}_{j_3}
\cdots \delta^{i_n}_{j_1}$.  Each pair of indices $i_k,j_k$ corresponds to an
external gluon.} \label{fig:n-gluon_colorflow}
\end{figure}

\begin{figure}[!t]
\begin{center}
\vspace*{0cm} \hspace*{0cm} \epsfxsize=11cm \epsfbox{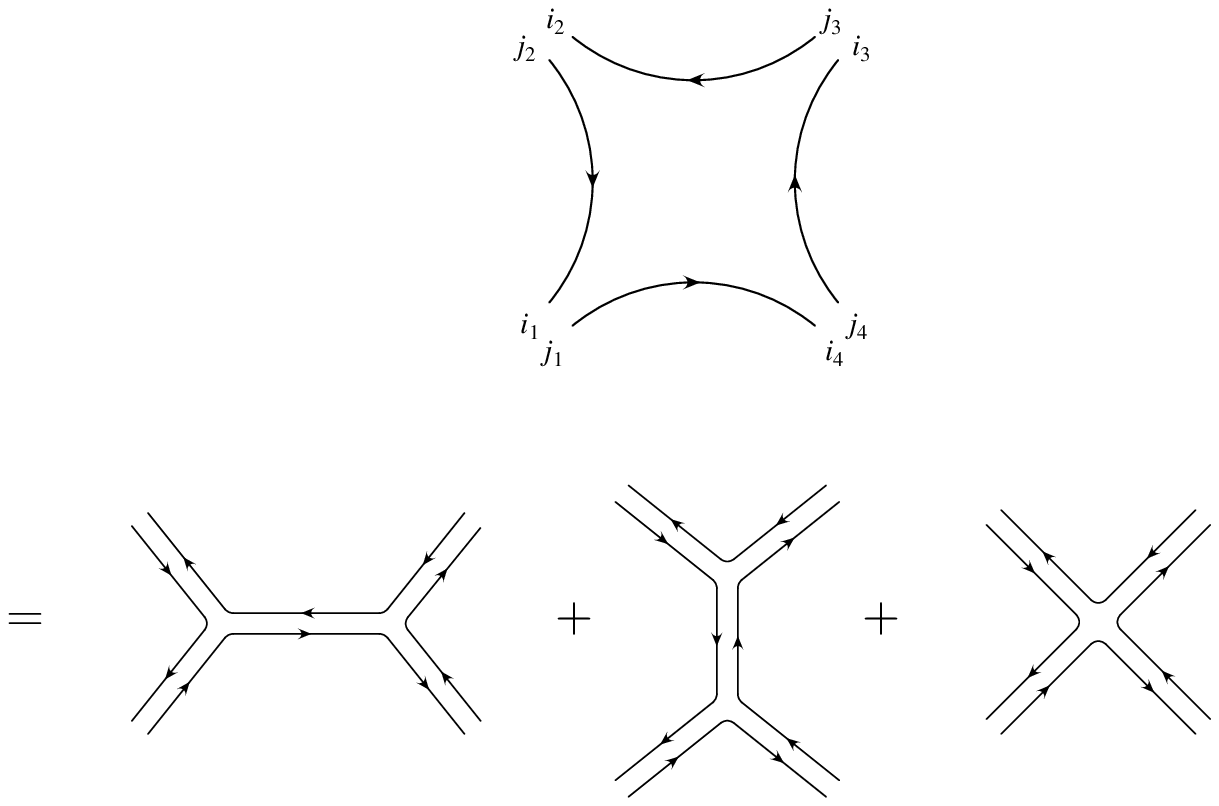}
\vspace*{0cm}
\end{center}
\caption{Feynman diagrams corresponding to a four-gluon partial amplitude.}
\label{fig:4-gluon_diagrams}
\end{figure}

To calculate $A(1,2,\ldots,n)$, one orders the gluons clockwise, as shown in
Fig.~\ref{fig:n-gluon_colorflow}, and draws color-flow lines, with color
flowing counter-clockwise, connecting adjacent gluons.  One then deforms the
color-flow lines in all possible ways to form the Feynman diagrams that
contribute to this partial amplitude.  An example of a four-gluon partial
amplitude is given in Fig.~\ref{fig:4-gluon_diagrams}.  At each vertex, one
needs only a single color flow in the three- and four-gluon Feynman rules
given in Fig.~\ref{fig:colorflow_feynmanrules}.  Thus, when constructing a
partial amplitude, the sum over permutations in the three- and four-gluon
vertices may be dropped.

It is evident that the Feynman diagrams that contribute to a partial amplitude
are planar. This is not due to an expansion in $1/N$; the partial amplitudes
are exact. The number of Feynman diagrams that contribute to an $n$-gluon
partial amplitude is listed in Table~\ref{tab:ndiagrams}.  The number grows
approximately like $3.8^n$, rather than the factorial growth of the number of
Feynman diagrams that contribute to the full amplitude, approximately $(2n)!$.

\begin{table}[p]
\caption{Number of Feynman diagrams contributing to an $n$-gluon partial
amplitude.  The number grows approximately like $3.8^n$. In contrast, the
number of Feynman diagrams contributing to the full amplitude grows
factorially, approximately $(2n)!$.}
\medskip
\addtolength{\arraycolsep}{0.1cm}
\renewcommand{\arraystretch}{1.4}
\begin{center} \begin{tabular}[4]{c|cc}
\hline \hline
& \multicolumn{2}{c}{\# diagrams} \\[1pt]
$n$ & partial amplitude & full amplitude
\\
\hline
4 &3     &4         \\[7pt]
5 &10    &25        \\[7pt]
6 &36    &220       \\[7pt]
7 &133   &2485      \\[7pt]
8 &501   &34300     \\[7pt]
9 &1991  &559405    \\[7pt]
10&7335  &10525900  \\[7pt]
11&28199 &224449225 \\[7pt]
12&108281&5348843500\\[7pt]
\hline \hline
\end{tabular}
\end{center}
\label{tab:ndiagrams}
\end{table}

This procedure is analogous to the ``color-ordered'' Feynman rules that have
been developed for the fundamental-representation decomposition of multi-gluon
amplitudes, Eq.~(\ref{eq:fund_decomposition})
\cite{Mangano:1990by,Mangano:1987xk}. The similarity of this procedure with
the construction of string amplitudes has been noted. Each gluon corresponds
to an open string with color-anticolor charges $i,j$ on its ends.  The diagram
in Fig.~\ref{fig:n-gluon_colorflow} represents the scattering of $n$ open
strings. The Feynman diagrams in Fig.~\ref{fig:4-gluon_diagrams} correspond to
the zero-slope limit of the scattering of four open strings
\cite{Mangano:1990by,Mangano:1987xk}.

One obtains the cross section from the $n$-gluon amplitude by squaring and
summing over the colors of the external gluons.  Since each external gluon has
two indices, naively summing counts $N^{2}$ colors per gluon.  To sum over
only the $N^2-1$ desired colors, it is sufficient to first apply the
projection operator
\begin{equation}
\label{eq:gluon_projector}
P^{i i'}_{j j'} \equiv \D{i}{j'} \D{i'}{j} - \frac{1}{N} \D{i}{j} \D{i'}{j'}
\end{equation}
to each external gluon before squaring and summing over colors.  However, the
second term in the projection operator corresponds to a $U(1)$ gluon, so it
does not couple to the $n$-gluon amplitude.  Hence, in the case of the
$n$-gluon amplitude, it is sufficient to naively sum over colors.  This is not
the case if external quarks are present, as we discuss in the following
section.

When squaring the amplitude and summing over colors, the leading term in $1/N$
is given by the square of each color flow:
\begin{equation}
\D{i_1}{j_2} \D{i_2}{j_3} \cdots \D{i_n}{j_1} (\D{i_1}{j_2} \D{i_2}{j_3}
\cdots \D{i_n}{j_1})^{\dag} = \D{i_1}{j_2} \D{i_2}{j_3} \cdots \D{i_n}{j_1}
(\D{j_2}{i_1} \D{j_3}{i_2} \cdots \D{j_1}{i_n}) = N^n .
\end{equation}
Cross terms between different color flows yield monomials $N^{n-m}$, where $m$
is even.  This contrasts with the squaring and summing over colors in the
fundamental-representation decomposition, Eq.~(\ref{eq:fund_decomposition}).
There, each term obtained is a polynomial in $N$, rather than a monomial
\cite{Mangano:1990by}.\footnote{In the case of the $n$-gluon amplitude in the
fundamental-representation decomposition, only the leading term in the
polynomial in $N$ need be retained, as subleading terms are associated with
$U(1)$ gluons \cite{Dixon:1996wi}.} In this sense, squaring and summing over
colors is simpler in the color-flow decomposition,
Eq.~(\ref{eq:colorflow_decomposition}).

For processes with many external gluons, it is necessary to sum over colors
using Monte-Carlo techniques in order to produce a cross section with
sufficient speed to be useful in practice
\cite{Caravaglios:1998yr,Mangano:2001xp,Draggiotis:1998gr,Draggiotis:2002hm}.
The color-flow decomposition is well suited for such a calculation.  One
chooses, via Monte-Carlo methods, a particular color assignment for the
external gluons. This is accomplished by randomly selecting the colors of the
upper and lower indices. A necessary (but not sufficient) condition for a
nonvanishing color assignment (one which has at least one color flow) is that
the number of upper and lower indices of the color $R$ must be the same;
similarly for the colors $G$ and $B$.  In general, only a small fraction of
the $(n-1)!$ color flows contribute to a given color assignment, so it is
necessary to evaluate only a small subset of the partial amplitudes. This is
crucial, as most of the computational time is spent on calculating the partial
amplitudes.  We give in the third column of Table~\ref{tab:colorflows} the
average number of partial amplitudes that contribute to a given nonvanishing
color assignment. Although this number grows factorially with the number of
gluons, it grows approximately as $(n/3)!$ rather than $(2n)!$.

\begin{table}[p]
\caption{Average number of partial amplitudes, for $n$-gluon scattering, that
must be evaluated per nonvanishing color assignment in three different color
decompositions: the fundamental-representation decomposition (using both the
Gell-Mann matrices and the matrices used in Ref.~\cite{Caravaglios:1998yr}),
the color-flow decomposition, and the adjoint-representation decomposition.
The fundamental-representation decomposition is much more efficient using the
matrices of Ref.~\cite{Caravaglios:1998yr}. The color-flow decomposition is
much more efficient than the fundamental-representation decomposition,
especially when $n$ is large. The adjoint-representation decomposition is
almost as efficient as the color-flow decomposition, but requires the
multiplication of sparse $9\times 9$ matrices.}
\medskip
\addtolength{\arraycolsep}{0.1cm}
\renewcommand{\arraystretch}{1.4}
\begin{center} \begin{tabular}[4]{c|cccc}
\hline \hline
& \multicolumn{4}{c}{Decomposition} \\[1pt]
& \multicolumn{2}{c}{Fundamental} && \\
$n$ & Gell-Mann & Ref.~\cite{Caravaglios:1998yr} & Color-flow & Adjoint
\\
\hline
4 &4.83   &3.02   &1.28 &1.15\\[7pt]
5 &15.2   &7.26   &1.83 &1.52\\[7pt]
6 &56.5   &20.6   &3.21 &2.55\\[7pt]
7 &251    &68.0   &6.80 &5.53\\[7pt]
8 &1280   &254    &17.0 &15.8\\[7pt]
9 &7440   &1080   &48.7 &56.4\\[7pt]
10&47800  &4930   &158  &243 \\[7pt]
11&337000 &25500  &570  &1210\\[7pt]
12&2590000&148000 &2250 &6750\\[7pt]
\hline \hline
\end{tabular}
\end{center}
\label{tab:colorflows}
\end{table}

We have written a code to identify the color flows that contribute to a given
color assignment.\footnote{This code is available at
http://madgraph.physics.uiuc.edu.} Once the color flows are identified, one
must evaluate the corresponding partial amplitudes. The amplitude for a given
color assignment is the sum of these partial amplitudes, with unit
coefficients, as per Eq.~(\ref{eq:colorflow_decomposition}).  No matrix
multiplication is necessary to evaluate the color coefficients.

Let us compare the efficiency of this procedure with that of the
fundamental-represen\-ta\-tion decomposition,
Eq.~(\ref{eq:fund_decomposition}).  First we use the standard Gell-Mann
matrices (see the Appendix of Ref.~\cite{Barger:nn}) to evaluate the color
coefficients.  The average number of partial amplitudes that must be evaluated
per nonvanishing color assignment is given in the first column of
Table~\ref{tab:colorflows}.\footnote{The results for $n=11,12$ are difficult to
calculate, so we approximate them by extrapolating the results for $n\le 10$.}
It is evident that the fundamental-representation decomposition using the
Gell-Mann basis is far less efficient than the color-flow decomposition.

In Ref.~\cite{Caravaglios:1998yr} a particular basis for the
fundamental-representation matrices is chosen in order to minimize the average
number of traces of matrices that contribute to a given nonvanishing color
assignment. That basis is
\begin{eqnarray}
&&    \lambda^1 = \frac{1}{\sqrt{2}} \left( \begin{array}{ccc}
                           0 & 1 & 0 \\
                           0 & 0 & 0 \\
                           0 & 0 & 0 \end{array} \right) \quad,\quad
      \lambda^2 = \frac{1}{\sqrt{2}} \left( \begin{array}{ccc}
                           0 & 0 & 1 \\
                           0 & 0 & 0 \\
                           0 & 0 & 0 \end{array} \right) \quad,\quad
      \lambda^3 = \frac{1}{\sqrt{2}} \left( \begin{array}{ccc}
                           0 & 0 & 0 \\
                           1 & 0 & 0 \\
                           0 & 0 & 0 \end{array} \right) \nonumber \quad,\\
&&   \lambda^5 = \frac{1}{\sqrt{2}} \left( \begin{array}{ccc}
                           0 & 0 & 0 \\
                           0 & 0 & 1 \\
                           0 & 0 & 0 \end{array} \right) \quad,\quad
      \lambda^6 = \frac{1}{\sqrt{2}} \left( \begin{array}{ccc}
                           0 & 0 & 0 \\
                           0 & 0 & 0 \\
                           1 & 0 & 0 \end{array} \right) \quad,\quad
      \lambda^7 = \frac{1}{\sqrt{2}} \left( \begin{array}{ccc}
                           0 & 0 & 0 \\
                           0 & 0 & 0 \\
                           0 & 1 & 0 \end{array} \right) \nonumber \quad,\\
&&   \lambda^4 = \frac{1}{{2}} \left( \begin{array}{ccc}
                           1 & 0 & 0 \\
                           0 &-1 & 0 \\
                           0 & 0 & 0 \end{array} \right) \quad,\quad
      \lambda^8 = \frac{1}{\sqrt{12}} \left( \begin{array}{ccc}
                           1 & 0 & 0 \\
                           0 & 1 & 0 \\
                           0 & 0 &-2 \end{array} \right) \quad.\nonumber
\end{eqnarray}
Using this basis, we give in the second column of Table~\ref{tab:colorflows}
the average number of partial amplitudes that contribute to a given
nonvanishing color assignment.\footnote{These numbers agree, within Monte-Carlo
uncertainty, with those given in Ref.~\cite{Caravaglios:1998yr} for
$n=8,9,10$.} Although much less than the results using the Gell-Mann basis,
these numbers are significantly greater than in the color-flow decomposition,
especially when $n$ is large. Hence the color-flow decomposition is much more
efficient than the fundamental-representation decomposition.

The reason the fundamental-representation decomposition of
Ref.~\cite{Caravaglios:1998yr} is less efficient for evaluating color
coefficients is due to the matrices $\lambda^4$ and $\lambda^8$ above, which
have more than one nonvanishing element, and may therefore appear in many
traces. To avoid this, it is advantageous to replace these two matrices with
the three matrices
\begin{eqnarray}
&&    \lambda^4 = \frac{1}{\sqrt{2}} \left( \begin{array}{ccc}
                           1 & 0 & 0 \\
                           0 & 0 & 0 \\
                           0 & 0 & 0 \end{array} \right) \quad,\quad
      \lambda^8 = \frac{1}{\sqrt{2}} \left( \begin{array}{ccc}
                           0 & 0 & 0 \\
                           0 & 1 & 0 \\
                           0 & 0 & 0 \end{array} \right) \quad,\quad
      \lambda^9 = \frac{1}{\sqrt{2}} \left( \begin{array}{ccc}
                           0 & 0 & 0 \\
                           0 & 0 & 0 \\
                           0 & 0 & 1 \end{array} \right) \quad,\nonumber
\end{eqnarray}
thereby expanding to the fundamental representation of $U(3)$. In so doing, one
is including the $U(1)$ gluon; however, we know that this gluon decouples from
the $n$-gluon tree amplitude, so no error is being made. Using this expanded
basis of fundamental-representation matrices is equivalent to the color-flow
decomposition, Eq.~(\ref{eq:colorflow_decomposition}), because each matrix is
proportional to a product of Kronecker deltas
($(\lambda^1)^i_j=\delta^i_1\delta^2_j/\sqrt 2$,
$(\lambda^2)^i_j=\delta^i_1\delta^3_j/\sqrt 2$, {\it etc.}). However, the
color-flow decomposition leads to a faster evaluation of the color
coefficients, since no matrix multiplication is necessary, while the
multiplication of sparse $3\times 3$ matrices is required in the
fundamental-representation decomposition.

Another method for Monte-Carlo summation over color is used in
Ref.~\cite{Draggiotis:1998gr}.  Although the fundamental-representation
decomposition is used, it is converted to the color-flow decomposition before
the summation over color is performed.  This is achieved via
\begin{equation}
f^{a_1a_2a_3}(\lambda^{a_1})^{i_1}_{j_1}(\lambda^{a_2})^{i_2}_{j_2}
(\lambda^{a_3})^{i_3}_{j_3}=-i(\delta^{i_1}_{j_2}\delta^{i_2}_{j_3}\delta^{i_3}_{j_1}
-\delta^{i_1}_{j_3}\delta^{i_3}_{j_2}\delta^{i_2}_{j_1})\;.
\end{equation}
This paper goes on to promote color from a discrete to a continuous variable,
and performs a Monte-Carlo integration over color.  It is not clear that
anything is gained by making color a continuous variable, since a Monte-Carlo
summation over color is already possible when color is a discrete variable.

We next consider the summation over color in the adjoint-representation
decomposition, Eq.~(\ref{eq:adjoint_decomposition}).  This decomposition uses
only the $(n-2)!$ linearly-independent partial amplitudes, so it is
potentially more efficient than the color-flow decomposition.  Rather than
using the standard $SU(3)$ structure constants based on the Gell-Mann matrices
(see the Appendix of Ref.~\cite{Barger:nn}), we use a set based on the nine
$U(3)$ fundamental-representation matrices above, again exploiting the fact
that the $U(1)$ gluon decouples.  The structure constants in this basis are
antisymmetric in the first two indices only, and are given by
($[\lambda^a,\lambda^b]=if^{abc}\lambda^c$)
\begin{eqnarray}
&&f^{314} = f^{138} = f^{512} = f^{167} =  f^{141} = f^{811} = f^{235} =
f^{624} = f^{269} = f^{721} = f^{242} = f^{922} = f^{376} \nonumber \\
&&= f^{433} = f^{383} = f^{653} = f^{758} = f^{579} = f^{585} = f^{955} =
f^{466} = f^{696} = f^{877} = f^{797} = i/\sqrt 2\;, \nonumber
\end{eqnarray}
where all other structure constants, not related to the above by interchange of
the first two indices, vanish.  This set of structure constants leads to a much
faster evaluation of the color coefficients in
Eq.~(\ref{eq:adjoint_decomposition}) than the standard set of structure
constants based on the Gell-Mann matrices.  For $n$-gluon scattering, the
average number of partial amplitudes that must be evaluated per nonvanishing
color assignment is given in the fourth column of
Table~\ref{tab:colorflows}.\footnote{The adjoint-representation decomposition
yields fewer nonvanishing color assignments than the color-flow
decomposition.  For example, for $n=4$, the number of nonvanishing color
assignments is 73 in the adjoint-representation decomposition, 127 in the
color-flow decomposition.} It is comparable in efficiency to the color-flow
decomposition.  However, the color-flow decomposition leads to a much faster
evaluation of the color coefficients, since the multiplication of sparse
$9\times 9$ matrices is required in the adjoint-representation decomposition.

To demonstrate the utility of the color-flow decomposition, we calculate the
subprocess cross sections with 11 and 12 external gluons at tree level ($gg \to
9g$ and $gg\to 10g$), results that have not yet appeared in the literature.  We
employ the same cuts as Ref.~\cite{Caravaglios:1998yr},
\begin{equation}
p_{Ti}>60\;{\rm GeV}\;,\;\;|\eta_i|<2\;,\;\;\Delta R_{ij}>0.7\;,
\end{equation}
where the subprocess cross section with 10 external gluons ($gg \to 8g$) at
$\sqrt{\hat s}=1500$ GeV was presented, using $\alpha_S=0.12$ for illustrative
purposes.  We use the Berends-Giele recursion relations \cite{Berends:1987me}
to obtain the partial amplitudes,\footnote{The number of calculations that
must be performed to evaluate a partial amplitude with these recursion
relations grows only linearly, in contrast to the number of Feynman diagrams,
which grows exponentially, approximately $3.8^n$ (see
Table~\ref{tab:ndiagrams}).} and evaluate the basic currents upon which these
relations are based using HELAS \cite{Murayama:1992gi}.  We increase the
subprocess energy for $n=11,12$ to $\sqrt{\hat s}=2000,2500$ GeV,
respectively, to maintain a roughly constant fraction of generated events that
pass the cuts.  No effort is made to optimize the generation of the phase
space; our goal is to show that the use of the color-flow decomposition speeds
up the calculation so much (a factor of about 40 for $n=11$ and about $60$ for
$n=12$, see Table~\ref{tab:colorflows}) that one can obtain the $gg \to 9g$
and $gg\to 10g$ subprocess cross sections with a straightforward phase-space
generator such as RAMBO \cite{Kleiss:1985gy}. Using this procedure, we confirm
the $n=10$ result of Ref.~\cite{Caravaglios:1998yr}, and give the $n=11,12$
results in Table~\ref{tab:sigma}.\footnote{The code NGLUONS is available at
http://madgraph.physics.uiuc.edu.}

\begin{table}[t]
\caption{Subprocess cross section (pb) for $gg\to (n-2)g$ at subprocess energy
$\sqrt{\hat s}$ (GeV).}
\medskip
\addtolength{\arraycolsep}{0.1cm}
\renewcommand{\arraystretch}{1.4}
\begin{center} \begin{tabular}[3]{c|c|c}
\hline \hline
$n$ & $\sqrt{\hat s}$ (GeV) & $\hat\sigma(gg\to (n-2)g)$ (pb)
\\
\hline
10&1500 &$0.70\pm 0.04$ \\[7pt]
11&2000 &$0.30\pm 0.02$ \\[7pt]
12&2500 &$0.097\pm 0.006$ \\[7pt]
\hline \hline
\end{tabular}
\end{center}
\label{tab:sigma}
\end{table}

The color-flow decomposition nicely lends itself to merging with a
shower Monte Carlo, such as HERWIG
\cite{Corcella:2000bw,Corcella:2001wc} or Pythia
\cite{Sjostrand:2001yu}, which is based on the color flow of a
given hard-scattering subprocess.  A given color assignment
typically has several color flows that contribute.  One of these
color flows is randomly chosen to be associated with the event,
weighted by the square of the partial amplitude corresponding to
that color flow.  The weight does not include a color coefficient
(since it is unity), unlike the fundamental-representation
decomposition \cite{Caravaglios:1998yr}. The event is then evolved
with a shower Monte Carlo. This neglects the interference between
different color flows, but this interference is suppressed by a
power of $1/N^2$.  This is not a deficiency of the color-flow
decomposition, but rather is an inherent feature of the shower
Monte-Carlo approximation for soft radiation
\cite{Marchesini:1987cf}.

\section{$\bar qq$ and $n$ gluons}\label{sec:1quark}

\begin{figure}[b]
\begin{center}
\hspace*{0in} \epsfxsize=1.8in \epsfbox{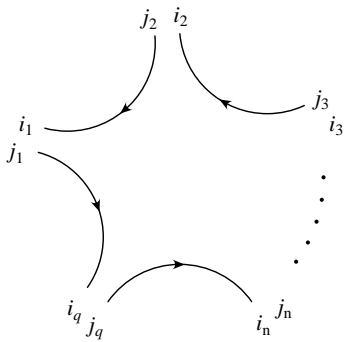}
\end{center}
\vspace*{-.1in} \caption{The color flow $\delta^{i_q}_{j_1} \delta^{i_1}_{j_2}
\cdots \delta^{i_n}_{j_q}$ for one $\bar qq$ pair and $n$ gluons.}
\label{fig:qq_colorflow}
\end{figure}

\begin{figure}[t!]
\begin{center}
\vspace*{0cm} \hspace*{0cm} \epsfxsize=11cm \epsfbox{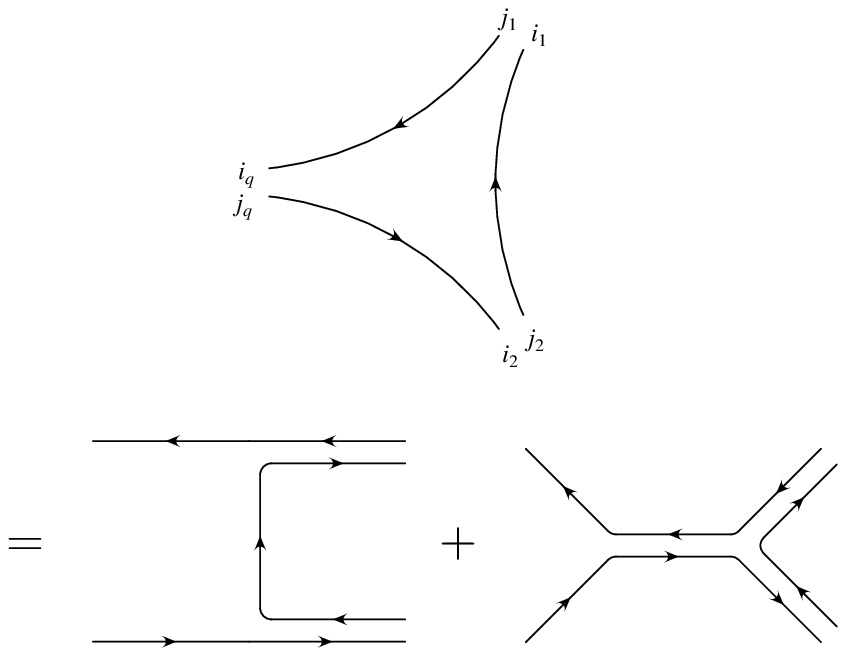}
\vspace*{0cm} \caption{Feynman diagrams corresponding to a partial
amplitude for one $\bar qq$ pair and two gluons.}
\label{fig:2g1q-cf}
\end{center}
\end{figure}

Consider the case where there is one quark line and $n$ gluons.  The outgoing
quark has color $i_q$, the outgoing antiquark has anticolor $j_q$.  The color
flow is identical to that of the $n$-gluon case, except the $\bar qq$ replaces
one of the gluons, as shown in Fig.~\ref{fig:qq_colorflow}.  The color-flow
decomposition is
\begin{equation}
{\cal M}(\bar qq+ng) = \sum_{P(1,\ldots,n)} \D{i_q}{j_1} \D{i_1}{j_2} \cdots
\D{i_n}{j_q} \; A(q,1,2,\ldots,n,\bar q)\;, \label{eq:qqdecomp}
\end{equation}
where the sum is over all $n!$ permutations of $(1,\ldots,n)$.  The arguments
$q$, $\bar q$ in the partial amplitude represent the momenta and helicities of
the outgoing quark and antiquark.  This decomposition follows directly from the
Feynman rules of Figs.~\ref{fig:colorflow_feynmanrules} and
\ref{fig:colorflow_gluon_propagator} and is similar to the decomposition of the
$n$-gluon amplitude.  The $U(1)$ gluon propagator, which couples only to
quarks, does not contribute at tree level since there is only one quark line.
As an example, the Feynman diagrams contributing to a particular partial
amplitude for the case of one $\bar qq$ and two gluons are shown in
Fig.~\ref{fig:2g1q-cf}.

Before squaring the amplitude and summing over colors, one must apply the
projection operator, Eq.~(\ref{eq:gluon_projector}), to each external gluon.
This generates terms proportional to powers of $(-1/N)$.  In the $n$-gluon
case, these terms vanish.  In the present case, they do not vanish due to the
presence of a quark line to which the external $U(1)$ gluon couples.  One
obtains
\begin{eqnarray}
P \cdots P{\cal M}(\bar qq+ng) & = & \sum_{P(1,\ldots,n)} \D{i_q}{j_1}
\D{i_1}{j_2} \cdots \D{i_n}{j_q} \; A(q,1,2,\ldots,n,\bar q)
\nonumber \\
& + & \left(-\frac{1}{N}\right) \sum_{P(1,\ldots,n)} \D{i_q}{j_1} \D{i_1}{j_2}
\cdots \D{i_{n-1}}{j_q}
\D{i_n}{j_n} \; A(q,1,2,\ldots,n-1,\bar q,n) \nonumber \\
& + & \left(-\frac{1}{N}\right)^2 \frac{1}{2!} \sum_{P(1,\ldots,n)}
\D{i_q}{j_1} \D{i_1}{j_2} \cdots \D{i_{n-2}}{j_q} \D{i_{n-1}}{j_{n-1}}
\D{i_n}{j_n} \;
A(q,1,2,\ldots,n-2,\bar q,n-1,n) \nonumber \\
& \vdots & \nonumber \\
& + & \left(-\frac{1}{N}\right)^n \D{i_q}{j_q} \D{i_1}{j_1} \cdots
\D{i_n}{j_n} \; A(q,\bar q,1,2,\ldots,n)\;, \label{eq:onequark}
\end{eqnarray}
where the partial amplitudes of the subleading terms in $1/N$ are linear
combinations of the leading partial amplitudes $A(q,1,2,\ldots,n,\bar q)$. The
subleading partial amplitude $A(q,1,\ldots,n-k,\bar q,n-k+1,\ldots,n)$
corresponds to the amplitude for $n-k$ gluons and $k$ $U(1)$ gluons.  For
example, the subleading partial amplitude for one $U(1)$ gluon is given by the
linear combination
\begin{equation}
A(q,1,2,\ldots,\bar q,n)=A(q,1,2,\ldots,n,\bar q)+A(q,1,2,\ldots,n,n-1,\bar
q)+\cdots+A(q,n,1,2,\ldots,\bar q)\;. \label{eq:onequarkward}
\end{equation}
This is analogous to the photon-decoupling equation\footnote{Also known as the
dual Ward identity.} for the $n$-gluon amplitude
\cite{Mangano:1990by,Mangano:1987xk,Bern:1990ux},
\begin{equation}
0=A(1,2,\ldots,n)+A(1,2,\ldots,n,n-1)+\cdots+A(1,n,2,\ldots,n-1)\;.
\end{equation}
More generally, the linear relations for the subleading partial amplitudes
with $k$ $U(1)$ gluons in terms of the leading partial amplitudes are
analogous to the Kleiss-Kuijf relations amongst the multi-gluon amplitudes
\cite{Kleiss:1988ne}.

The sum over permutations of the subleading terms with $k$ $U(1)$ gluons
contains a factor $1/k!$, because terms that differ only by the exchange of
$U(1)$ gluons are identical.  Thus there are $n!/k!$ different permutations in
the terms with $k$ $U(1)$ gluons.   There is only a single term in which all
gluons are $U(1)$, given at the end of Eq.~(\ref{eq:onequark}).

\begin{figure}[t!]
\vspace*{0cm} \hspace*{0cm} \epsfxsize=20cm \epsfbox{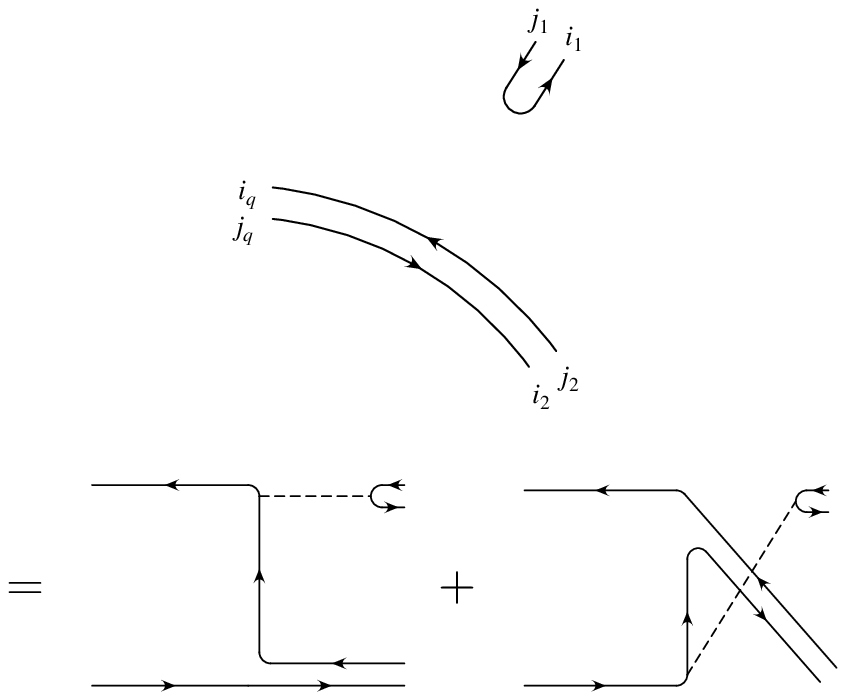}
\caption{Feynman diagrams corresponding to a partial amplitude for one $\bar q
q$ pair, a gluon, and a U(1) gluon.} \label{fig:2g1q-cf-u1}
\end{figure}

It is more efficient to calculate the subleading partial amplitudes directly,
rather than as a linear combination of the leading partial amplitudes.  This
is done by replacing $k$ of the external gluons by $U(1)$ gluons, and
associating a factor $(-1/N)$ with each $U(1)$ gluon. For example, the Feynman
diagrams for the a subleading partial amplitude for the case of a $\bar qq$
pair, one gluon, and one $U(1)$ gluon are shown in Fig.~\ref{fig:2g1q-cf-u1}.
Since the $U(1)$ gluon couples only to quarks, the non-Abelian diagram present
in the leading partial amplitude, Fig.~\ref{fig:2g1q-cf}, does not appear. The
non-Abelian diagram cancels when the subleading partial amplitude is
calculated as a linear combination of the leading partial amplitudes via
Eq.~(\ref{eq:onequarkward}).  Another feature of subleading partial amplitudes
is that there are contributions from nonplanar diagrams, since a $U(1)$ gluon
can be attached to any quark line without changing the color flow.  Thus the
number of Feynman diagrams contributing to a subleading partial amplitude with
$k$ $U(1)$ gluons grows like $k!$.\footnote{Using the Berends-Giele recursion
relations \cite{Berends:1987me}, the number of computations necessary to
evaluate such a partial amplitude grows only exponentially rather than
factorially.}

The amplitude of Eq.~(\ref{eq:onequark}) can be used to perform a Monte-Carlo
summation over color. As in the all-gluon case, one first selects a color
assignment by randomly choosing the colors of the upper and lower indices,
then checking that the number of upper and lower $R$ indices are the same, and
similarly for $B$ and $G$; this is a necessary (but not sufficient) condition
for a nonvanishing color assignment. One identifies the color flows
corresponding to this color assignment, including the subleading color flows.
The partial amplitudes corresponding to these color flows are then evaluated.
The amplitude is the sum of the partial amplitudes, with coefficients of
$(-1/N)$ raised to the power of the number of $U(1)$ gluons, as per
Eq.~(\ref{eq:onequark}).

In Table~\ref{tab:1qcolorflows} we compare the efficiency of the color-flow
decomposition with that of the fundamental-representation decomposition, which
is given by \cite{Mangano:1990by,Kunszt:1985mg,Mangano:1987kp,Mangano:1988kk}
\begin{equation}
{\cal M}(\bar qq+ng) = \sum_{P(1,\ldots, n)} (\lambda^{a_1} \cdots
\lambda^{a_n})^{i_q}_{j_q} \; A(q,1,\ldots,n,\bar q)\;.
\label{eq:1qfund_decomposition}
\end{equation}
In the fundamental-representation decomposition, the matrices of
Ref.~\cite{Caravaglios:1998yr}, given in the previous section, are used.  We
list the average number of partial amplitudes, both leading and subleading,
that must be evaluated per nonvanishing color assignment.  As in the all-gluon
case, the color-flow decomposition is much more efficient when the number of
external gluons is large.  The gain is not as large as in the all-gluon case,
however. This is to be expected, as external quarks are treated identically in
the color-flow and fundamental-representation decompositions; only the gluons
are treated differently \cite{Mangano:1988kk}.  The color-flow decomposition
is of comparable efficiency for the case of one $\bar qq$ pair and no $\bar qq$
pairs (Table~\ref{tab:colorflows}) for a given number of external particles.

\begin{table}[p]
\caption{Average number of partial amplitudes, for the scattering of one $\bar
qq$ pair and $n$ gluons, that must be evaluated per nonvanishing color
assignment in two different color decompositions: the
fundamental-representation decomposition used in
Ref.~\cite{Caravaglios:1998yr}, and the color-flow decomposition. The
color-flow decomposition is much more efficient than the
fundamental-representation decomposition, especially when $n$ is large.}
\medskip
\addtolength{\arraycolsep}{0.1cm}
\renewcommand{\arraystretch}{1.4}
\begin{center} \begin{tabular}[3]{c|cc}
\hline \hline
& \multicolumn{2}{c}{Decomposition} \\[1pt]
$n$ & Fund.~(Ref.~\cite{Caravaglios:1998yr}) & Color-flow
\\
\hline
2 &1.44 &1.55\\[7pt]
3 &2.56 &2.22\\[7pt]
4 &5.66 &3.66\\[7pt]
5 &15.3 &7.14\\[7pt]
6 &48.8 &16.3\\[7pt]
7 &179  &42.6\\[7pt]
8 &748  &126 \\[7pt]
9 &3460 &417 \\[7pt]
10&17400&1520\\[7pt]
\hline \hline
\end{tabular}
\end{center}
\label{tab:1qcolorflows}
\end{table}

In the previous section, we showed that the fundamental-representation
decomposition is equivalent to the color-flow decomposition for the $n$-gluon
amplitude when the fundamental representation is expanded to include a ninth
matrix.  This basis of nine matrices includes the $U(1)$ gluon, but since this
particle decouples from the $n$-gluon amplitude, no error is being made.  This
same procedure cannot be carried out for amplitudes involving quarks, since the
$U(1)$ gluon couples to quarks.  Terms must be added to cancel the contribution
of the $U(1)$ gluons; this is the role of the subleading terms in the
color-flow decomposition, Eq.~(\ref{eq:onequark}).

Merging the hard-scattering cross section for $\bar qq + ng$ with a shower
Monte Carlo program proceeds similarly as in the case of all gluons discussed
in the previous section.  For a given color assignment, one weights each
contributing color flow by the square of the partial amplitude (including the
square of the corresponding power of $(-1/N)$).  The color flow associated
with the event is then randomly selected from the weighted color flows
\cite{Mangano:2001xp}.

\section{Two $\bar qq$ and $n$ gluons}\label{sec:2quark}

\begin{figure}[t!]
\vspace*{0cm} \hspace*{0.5cm} \epsfxsize=15cm \epsfbox{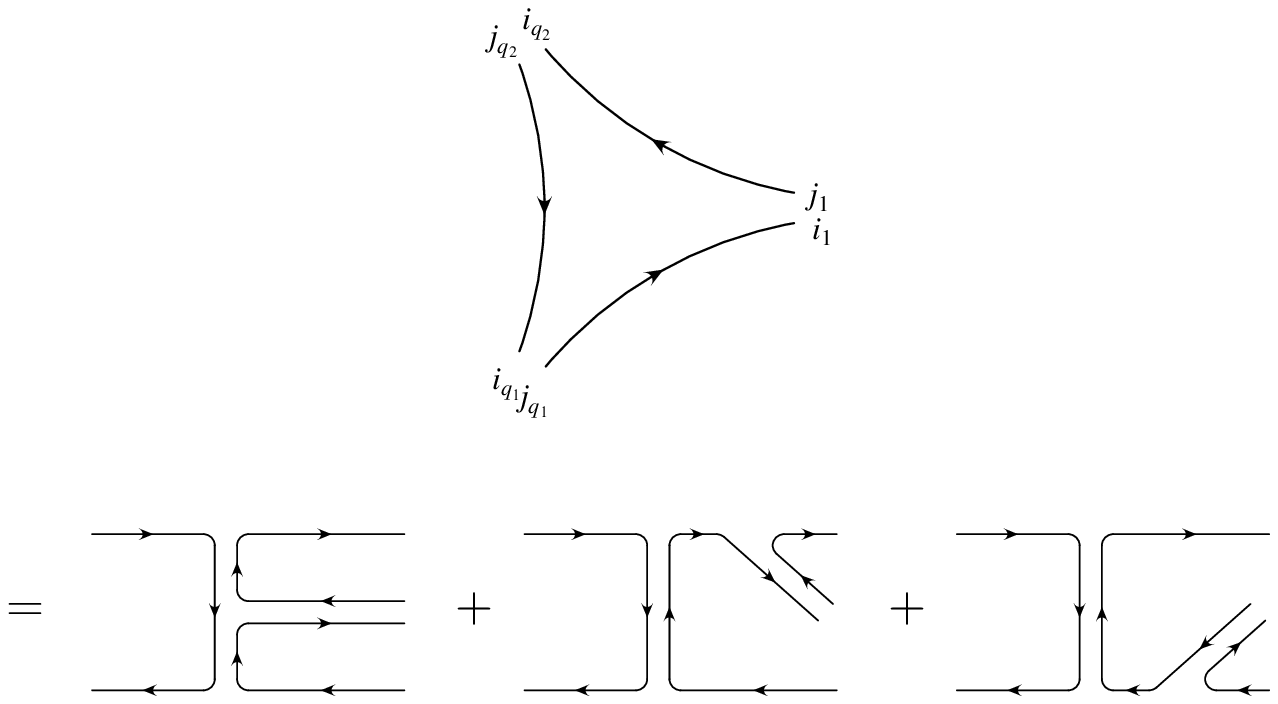}
\caption{Feynman diagrams corresponding to a partial amplitude for two
(distinguishable) $\bar qq$ pairs and a gluon.} \label{fig:1g2q-cf}
\end{figure}

We now consider two quark pairs and $n$ gluons.  Since there are two quark
lines, a new feature enters: Feynman diagrams with a $U(1)$ gluon exchanged
between the quark lines.  These diagrams are suppressed by $1/N$ due to the
propagator of the $U(1)$ gluon (see Fig.~\ref{fig:colorflow_gluon_propagator}).

\begin{figure}[p]
\begin{center}
\vspace*{0cm} \hspace*{0.5cm} \epsfxsize=10cm \epsfbox{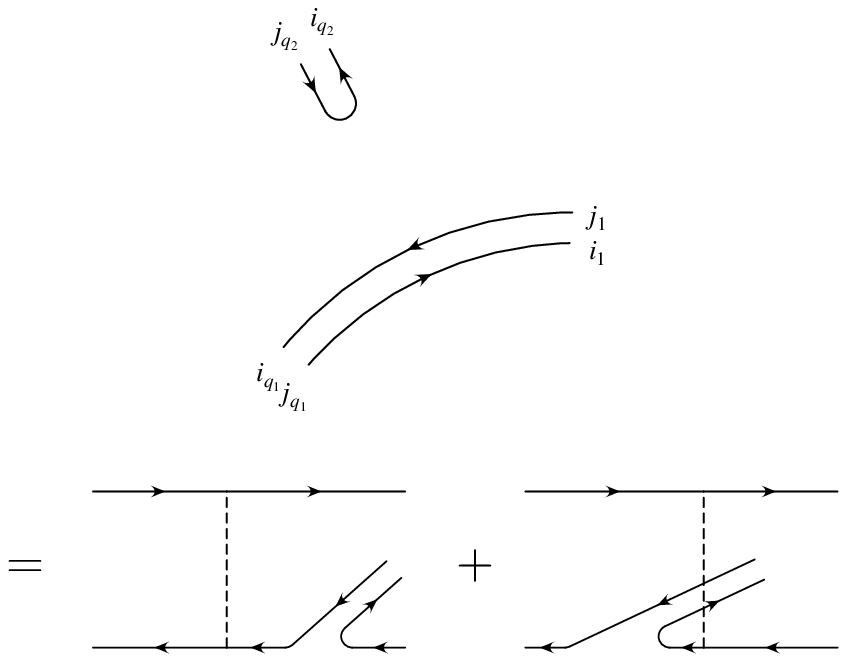}
\end{center}
\caption{Feynman diagrams corresponding to a subleading partial amplitude for
two (distinguishable) $\bar qq$ pairs and a gluon.} \label{fig:1g2q-cf-u1}
\end{figure}

The leading partial amplitudes in $1/N$ have a color flow
analogous to that of Fig.~\ref{fig:qq_colorflow}, but with two
$\bar qq$ pairs.  For example, we show in Fig.~\ref{fig:1g2q-cf}
the Feynman diagrams contributing to a partial amplitude for two
(distinguishable) quark pairs and one gluon.  In contrast, we show
in Fig.~\ref{fig:1g2q-cf-u1} the Feynman diagrams contributing to
a subleading partial amplitude, which contains an internal $U(1)$
gluon. Because the $U(1)$ gluon carries no color, the color flow
factors into two separate color flows, each beginning with a quark
and ending with an antiquark. The color-flow decomposition for 2
$\bar qq$ pairs and $n$ gluons is thus
\begin{eqnarray}
&&{\cal M}(\bar q_1q_1+\bar q_2q_2 +ng) = \sum_{P(q_2,1,\ldots,n)}
\D{i_{q_1}}{j_{q_2}}\D{i_{q_2}}{j_1} \D{i_1}{j_2} \cdots
\D{i_n}{j_{q_1}} \; A(q_1,\bar q_2,q_2, 1,2,\ldots,n,\bar q_1)
\nonumber
\\
&&-\frac{1}{N}\sum_{P(1,\ldots,n)}\sum_{r=0}^{n} (\D{i_{q_1}}{j_1}
\D{i_1}{j_2} \cdots \D{i_r}{j_{q_1}}) \; (\D{i_{q_2}}{j_{r+1}}
\D{i_{r+1}}{j_{r+2}} \cdots \D{i_n}{j_{q_2}}) \;
A(q_1,1,2,\ldots,r,\bar q_1,q_2,r+1,\ldots,n,\bar q_2)\;. \nonumber\\
\label{eq:twoquarks}\end{eqnarray}
The second line contains the subleading terms, with the factor
$-1/N$ from the $U(1)$-gluon propagator made explicit.  A sum over
permutations of a sum over partitions of the gluons between the
two quark lines is performed on these terms.

\begin{figure}[p]
\begin{center}
\vspace*{0cm} \hspace*{0.5cm} \epsfxsize=10cm \epsfbox{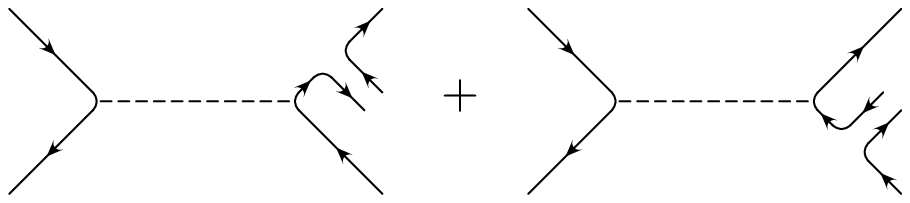}
\end{center}
\caption{Additional Feynman diagrams contributing to the partial amplitude of
Fig.~\ref{fig:1g2q-cf} for two identical $\bar qq$ pairs and a gluon.}
\label{fig:1g2q-cf-ident}
\end{figure}

The color-flow decomposition for two identical $\bar qq$ pairs is
similar. However, the first and second lines in
Eq.~(\ref{eq:twoquarks}) no longer correspond to leading and
subleading in $1/N$; both contain terms that are leading and
subleading.  In the example of two $\bar qq$ pairs and one gluon
given above, the additional (subleading) Feynman diagrams in
Fig.~\ref{fig:1g2q-cf-ident} contribute to the color flow in the
first line of Eq.~(\ref{eq:twoquarks}) if the quark pairs are
identical.

\begin{figure}[!t]
\begin{center}
\vspace*{0cm} \hspace*{0.5cm} \epsfxsize=10cm \epsfbox{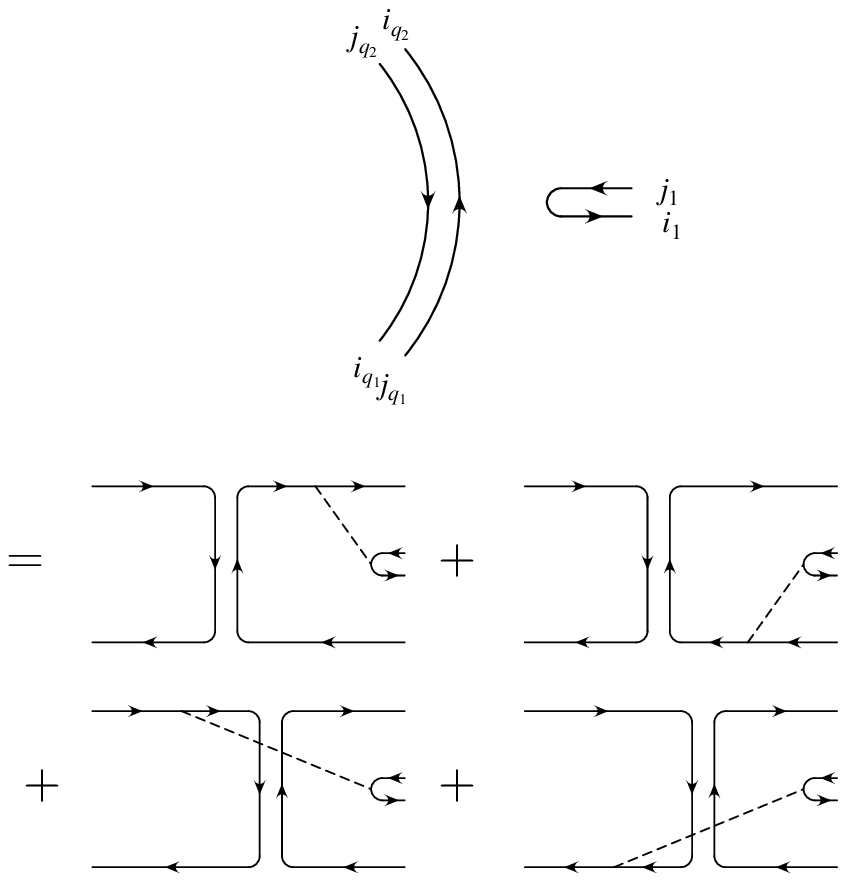}
\end{center}
\caption{Feynman diagrams corresponding to a partial amplitude for two
(distinguishable) $\bar qq$ pairs and a $U(1)$ gluon.} \label{fig:1p2q-cf}
\end{figure}

As usual, one must apply a projection operator,
Eq.~(\ref{eq:gluon_projector}), to each external gluon before squaring the
amplitude. As we saw in the previous section, this can be done
diagrammatically by replacing external gluons with $U(1)$ gluons.  For
example, in the case of two (distinguishable) $\bar qq$ pairs and one external
gluon, the Feynman diagrams contributing to a subleading term generated by
applying the projection operator to the external gluon are shown
diagrammatically in Fig.~\ref{fig:1p2q-cf}.

\section{General case}\label{sec:general}

The general case of any number of $\bar qq$ pairs and gluons follows the same
pattern established in the previous section.  Consider, for example, the case
of six $\bar qq$ pairs and any number of gluons.  A typical term in the
color-flow decomposition is
\begin{eqnarray}
&&(\D{i_{q_1}}{}\cdots\D{}{j_{q_2}}\D{i_{q_2}}{}\cdots\D{}{j_{q_3}}\D{i_{q_3}}{}
\cdots\D{}{j_{q_1}})
(\D{i_{q_4}}{}\cdots\D{}{j_{q_5}}\D{i_{q_5}}{}\cdots\D{}{j_{q_4}})
(\D{i_{q_6}}{}\cdots\D{}{j_{q_6}}) \nonumber\\
&&\hspace{2cm}\times A(q_1,\ldots,\bar q_2,q_2,\ldots,\bar q_3,q_3,\ldots,\bar
q_1,q_4,\ldots,\bar q_5,q_5,\ldots,\bar q_4,q_6,\ldots,\bar q_6)\;,
\end{eqnarray}
where the gluon labels are tacit.  If the quarks are distinguishable, this term
is of order $(-1/N)^2$, since there are three separate color flows joined by two
$U(1)$ gluons.  If some of the quarks are identical, this term does not correspond to a
unique order in $1/N$, as we saw in the previous section. As
usual, one must apply a projection operator,
Eq.~(\ref{eq:gluon_projector}), to each external gluon before
squaring the amplitude.

The color-flow decomposition differs from the fundamental-representation
decomposition in the treatment of the gluons.  Thus, in the case of an
amplitude with only external quarks and antiquarks, the two decompositions are
identical \cite{Mangano:1990by,Mangano:1988kk}.

\section{Conclusions}\label{sec:conclusions}

We have described a new color decomposition for tree-level multi-parton
amplitudes in an $SU(N)$ gauge theory. This decomposition is based on color
flow, which corresponds to the conservation of color in QCD.  An amplitude is
decomposed into a sum of partial amplitudes, each of which corresponds to a
particular color flow, and has a coefficient which is a power of $(-1/N)$.
These partial amplitudes are constructed from the color-flow Feynman rules of
Figs.~\ref{fig:colorflow_feynmanrules} and
\ref{fig:colorflow_gluon_propagator}.  The color-flow decomposition is a very
natural way to organize a calculation of a multi-parton amplitude.  Although
we have discussed the color-flow decomposition at tree level, it is clear from
the Feynman rules that it may be applied at the loop level as well
\cite{Dixon:1996wi,Bern:1990ux}.

The color-flow decomposition of a multi-parton amplitude is free of
fundamental-represen\-ta\-tion matrices and structure constants --- they simply
never occur in the construction of the amplitude.  This allows for a very
efficient numerical evaluation of the amplitude using Monte-Carlo techniques.
We showed that multi-parton amplitudes may be evaluated much more efficiently
in the color-flow decomposition than in the fundamental-representation
decomposition that has traditionally been used for such calculations.  This
will lead to faster codes for the calculation of multi-jet processes in QCD,
which are the dominant backgrounds to signals for new physics at hadron and
$e^+e^-$ colliders.  The color-flow decomposition also lends itself nicely to
the merging of the hard-scattering cross section with shower Monte-Carlo
programs that use the color flow to evolve parton final states into jets of
hadrons.

The color-flow decomposition applies not only to pure QCD processes, but also
to processes with additional particles, such as leptons, photons, $W$ and $Z$
bosons, and the Higgs boson.  The color-flow decomposition will be implemented
in the new event generator MadEvent \cite{Maltoni:2002qb}, based on the code
MadGraph \cite{Stelzer:1994ta}.

\section*{Acknowledgments}

\indent\indent This work was supported in part by the U.~S.~Department of
Energy under contract Nos.~DE-FG02-91ER40677 and DE-AC02-76CH03000.

\section*{Appendix}

In this appendix we prove that the partial amplitudes that appear in the
color-flow decomposition of the $n$-gluon amplitude,
Eq.~(\ref{eq:colorflow_decomposition}), are identical to those that appear in
the fundamental-representation decomposition,
Eq.~(\ref{eq:fund_decomposition}).  We then show that this is also true in the
general case of one or more $\bar qq$ pairs and any number of gluons.

The connection between the two decompositions is made by contracting each of
the matrices in the fundamental-representation decomposition, $\lambda^{a}$,
with the matrix ($\lambda^{a})^i_j$, and using
\begin{equation}
(\lambda^{a})^i_j(\lambda^{a})^{i'}_{j'}=\delta^{i}_{j'}\delta^{i'}_{j}-\frac{1}{N}
\delta^{i}_{j}\delta^{i'}_{j'}\;. \label{eq:deltadelta}
\end{equation}
The color coefficient of the fundamental-representation decomposition is
thereby transformed into that of the color-flow decomposition, plus additional
terms suppressed by powers of $1/N$:
\begin{equation}
{\rm Tr}\, (\lambda^{a_1}\lambda^{a_2}\cdots\lambda^{a_n})\;
(\lambda^{a_1})^{i_1}_{j_1}(\lambda^{a_2})^{i_2}_{j_2}\cdots(\lambda^{a_n})^{i_n}_{j_n}
=\delta^{i_1}_{j_2}\delta^{i_2}_{j_3}\cdots\delta^{i_n}_{j_1} + {\cal
O}(1/N)\;.\label{eq:projection}
\end{equation}
If we ignore the additional terms of ${\cal O}(1/N)$, we can conclude that the
partial amplitudes in the two decompositions are identical.  We now prove that
these additional terms do indeed cancel.

We know on physical grounds that the ${\cal O}(1/N)$ terms, which correspond
to $U(1)$ gluons, cancel in the full amplitude.  This can be proven via the
Kleiss-Kuijf relations amongst the partial amplitudes \cite{Kleiss:1988ne}.
Here we present a simpler proof, which uses the adjoint-representation
decomposition, Eq.~(\ref{eq:adjoint_decomposition}), as an intermediate step.

Using the Kleiss-Kuijf relations, it was shown in Ref.~\cite{DelDuca:1999rs}
that the partial amplitudes in the fundamental-representation decomposition are
equal to those in the adjoint-representation decomposition. We therefore need
only show that the ${\cal O}(1/N)$ terms vanish when we relate the
adjoint-representation decomposition to the color-flow decomposition.  The
color coefficients in the adjoint-representation decomposition may be written
\begin{equation}
(F^{a_2} F^{a_3} \cdots F^{a_{n-1}})^{a_1}_{a_n}={\rm Tr}\, (
\lambda^{a_1}[\lambda^{a_2},\ldots,[\lambda^{a_{n-1}},\lambda^{a_n}]\ldots])\;,
\label{eq:adjointtrace}
\end{equation}
using $[\lambda^a,\lambda^b]=if^{abc}\lambda^c$.  We now contract this with
$(\lambda^{a_1})^{i_1}_{j_1}\cdots (\lambda^{a_n})^{i_n}_{j_n}$ to transform to
the color-flow decomposition.

Using Eq.~(\ref{eq:deltadelta}), it is easy to show that for an arbitrary
$N\times N$ matrix $M$,
\begin{equation}
[\lambda^a,M]^i_j (\lambda^a)^{i'}_{j'} =
\delta^i_{j'}M^{i'}_j-M^i_{j'}\delta^{i'}_j\;,
\end{equation}
where the $1/N$ terms in Eq.~(\ref{eq:deltadelta}) have cancelled.  Similarly,
for arbitrary $N\times N$ matrices $M,O$,
\begin{equation}
{\rm Tr}\, (\lambda^a [M,O]) (\lambda^a)^i_j = [M,O]^i_j\;,
\end{equation}
where the $1/N$ terms have again cancelled.  Applying these two relations to
the contraction of Eq.~(\ref{eq:adjointtrace}) with
$(\lambda^{a_1})^{i_1}_{j_1}\cdots (\lambda^{a_n})^{i_n}_{j_n}$ shows that the
only terms that survive are of the form of the first term on the right-hand
side of Eq.~(\ref{eq:projection}); all $1/N$ terms cancel. This completes the
proof that the partial amplitudes in the color-flow decomposition,
Eq.~(\ref{eq:colorflow_decomposition}), are the same as in the
fundamental-representation decomposition, Eq.~(\ref{eq:fund_decomposition}).

In the general case of one or more $\bar qq$ pairs and $n$ gluons, one again
transforms from the fundamental-representation decomposition to the color-flow
decomposition by contracting with $(\lambda^{a_1})^{i_1}_{j_1}\cdots
(\lambda^{a_n})^{i_n}_{j_n}$ and applying Eq.~(\ref{eq:deltadelta}).  The $1/N$
terms that are generated are just the $1/N$ terms obtained in the color-flow
decomposition by applying the projection operator,
Eq.~(\ref{eq:gluon_projector}), to the external gluons.


\end{document}